\documentclass[12pt,a4wide,journal,onecolumn]{IEEEtran}
\usepackage{cite} 
\usepackage{url}  
\usepackage{ifthen}  
\usepackage{multicol}   
\usepackage{psfig}
\usepackage{epsfig}
\usepackage{subfigure}
\usepackage{amsopn,amsmath,amssymb,amsfonts}
\usepackage{graphicx}
\usepackage{graphics}
\usepackage{array}
\usepackage{multirow}
\usepackage{psbox}
\usepackage{cite}
\usepackage{citesort}
\usepackage{ifpdf}
\usepackage{epstopdf}
\usepackage{mdwmath}
\usepackage{mdwtab}
\usepackage{amsmath}
\usepackage{graphicx}
\usepackage{amssymb}
\newcommand{\qed}{\hfill \mbox{\raggedright \rule{.07in}{.1in}}}
\usepackage{balance}
\usepackage{float}
\newfloat{longequation}{t}{ext}

\newtheorem{thm}{Theorem}[section]

\newtheorem{rmk}{Remark}[section]

\begin{document}

\title{
Eigenvalue Based Sensing and SNR Estimation for
Cognitive Radio in Presence of Noise Correlation}
\author{
Shree Krishna Sharma, Symeon Chatzinotas, and Bj$\ddot{\mathrm{o}}$rn Ottersten \footnote{\textbf{This work has been submitted to the IEEE for possible publication. Copyright may be transferred without notice, after which this version may no longer be accessible.}}
\thanks{The authors are with the Interdisciplinary Centre for Security, Reliability and Trust (SnT), University of Luxembourg (http://www.securityandtrust.lu) email: \{shree.sharma, symeon.chatzinotas, bjorn.ottersten\}@uni.lu.}  }
\maketitle
\bigskip
\bigskip
\begin{abstract}
\baselineskip20pt
Herein, we present a detailed analysis of an eigenvalue based sensing technique in the presence of correlated noise in the context of a Cognitive Radio (CR). We use a Standard Condition Number (SCN) based decision statistic based on asymptotic Random Matrix Theory (RMT) for decision process. Firstly, the effect of noise correlation on eigenvalue based Spectrum Sensing (SS) is studied analytically under both the noise only and the signal plus noise hypotheses. Secondly, new bounds for the SCN are proposed for achieving improved sensing in correlated noise scenarios. Thirdly, the performance of Fractional Sampling (FS) based SS is studied and a method for determining the operating point for the FS rate in terms of sensing performance and complexity is suggested. Finally, an SNR estimation technique based on the maximum eigenvalue of the received signal's covariance matrix is proposed. It is shown that proposed SCN-based threshold improves sensing performance in the presence of correlated noise and SNRs upto $0$ dB can be reliably estimated without the knowledge of noise variance.
\end{abstract}
\baselineskip20pt
\bigskip
\newpage
\section{Introduction}
Spectrum Sensing (SS) plays an important role in Cognitive Radio (CR) networks in order to acquire the spectrum awareness required by CRs. The three main signal processing techniques for sensing the presence
of a primary user (PU) that appear in the literature are matched filter detection, Energy Detection (ED) and cyclostationary feature detection \cite{Aja:09}. Matched filter detection and cyclostationary feature detection techniques require the prior knowledge of the PU's signal to make the decision about the presence or absence of the PU signal \cite{Yar:09}. Although ED technique does not require any prior knowledge of PU's signal, the performance of
this technique is susceptible to noise covariance uncertainty \cite{Tsa:08}. Since both the prior knowledge about the PU's signal and the noise variance are unknown to the CRs in practical scenarios,
exploring efficient and blind SS techniques for CRs has emerged as an important research challenge. Several blind SS techniques have been proposed in the literature  \cite{Lhw:11,Osn:11,Bns:11,Dpy:08} without
requiring the prior knowledge of the PU's signal, the channel and the noise power. Furthermore, the performance of traditional SS techniques is limited by received signal
strength which may be severely degraded in multi-path fading and shadowing environments. Different diversity enhancing techniques such as multi-antenna, cooperative and oversampled techniques have been introduced in the literature to enhance the SS efficiency in wireless
fading channels \cite{Yyl:09,Wzha:12,Krs:11}. Most of these methods use the properties of the eigenvalues of the received signal's covariance matrix and use recent results from advances in Random Matrix Theory (RMT) \cite{Ant:04,Rmd:11}. The main advantage of eigenvalue based SS over other SS techniques is that it does not require any prior information of the PU's signal and it outperforms ED techniques, especially in the presence of noise covariance uncertainty \cite{Yyl:09}.

In this paper, we use the standard condition number (SCN) of the noise covariance matrix to analyze the effect of noise correlation on eigenvalue based SS technique. The SCN of a matrix is defined as the ratio of the maximum eigenvalue to the minimum eigenvalue \cite{Mmm:10} and can be used as a metric to characterize the support of the asymptotic eigenvalue probability distribution function (a.e.p.d.f.) of a random matrix. Furthermore, we use the SCN of the received signal's covariance matrix for decision process. If the calculated SCN is greater than noise only SCN, the decision is that a PU signal is present. Since noise correlation affects the SCN of the noise covariance matrix and as a result, the SCN of the received signal's covariance matrix, the decision metric is also affected. According to author's knowledge, this method has not been considered in the literature for SS in the presence of noise correlation.

Several blind SS techniques utilizing various features of the received signal's covariance matrix such as statistical covariance \cite{Yyc:09}, autocorrelation \cite{Nit:10} and eigenvalue distribution \cite{Yyl:09}
have been proposed in the literature. In most of the existing eigenvalue based SS literature, the authors consider asymptotically  large matrices whose eigenvalues are known to follow the
Marchenko-Patur (MP) law, which establishes the convergence  of the largest and smallest eigenvalues of these matrices. The authors in \cite{Cdb:08}  use this MP law to test a binary hypothesis under
white noise conditions using SCN for Wishart matrices. However, the sample covariance matrix of the noise is not a Wishart random matrix in the presence of correlated noise \cite{Yyl:09}. In practical situations, noise correlation arises due to oversampling and imperfections in filtering \cite{Yyl:09}. For example, when a narrowband signal is sampled with a sampling rate greater than the symbol rate, the noise sequence is not white. In case of noise correlation, the eigenvalue distribution does not follow the MP law and the SCN threshold proposed in \cite{Cdb:08} may result in degraded PU sensing performance. Therefore, new SCN-based sensing thresholds need to be investigated for carrying out SS in the presence of noise correlation. This is the first contribution of this paper.

Furthermore, several methods based on eigenvalue distribution of received signal's covariance matrix usually focus on interweave CR meaning that Secondary User (SU) transmits only when no PU signal is present \cite{Yyl:09,Wzha:12}. However, if side information is available about the primary Signal to Noise Ratio (SNR), advanced underlay transmission schemes could be employed at CRs. In practical scenarios, it would be advantageous to estimate the primary SNR in order to decide the transmission strategy of the cognitive transmitter. Depending on the estimated primary SNR level, different underlay transmission strategies (e.g. cognitive resource allocation) can be implemented at cognitive transmitter to allow the coexistence of primary and secondary systems. In this direction, we derive the a.e.p.d.f. of received signal's covariance matrix for signal plus noise case under white and correlated noise. The a.e.p.d.f. is then used to determine the maximum eigenvalue which is in turn exploited to estimate the SNR. Moreover, the SNR estimation performance is evaluated based on normalized mean square error (MSE). This is the second contribution of this paper.

The sampling rate in the receiver can be increased beyond the symbol rate, known as fractional sampling (FS), to enhance SS efficiency under fading channel conditions. FS is commonly used to enhance signal detection reliability in the receiver \cite{Tcc:04,Isa:09,Nts:07}. From the CR point of view, an FS receiver can be modeled as a virtual multiple-output system with presumably independent channel fading effects. This technique is especially beneficial in time varying channels with large Doppler spread, i.e. small channel coherence time. Another motivation for introducing the FS concept in the context of CR is that using more antennas at the receive-side is often impractical and expensive requiring multiple RF chains. In wireless fading environments, FS introduces diversity and can improve signal detection. However, FS also results in colored noise when the bandwidth of receive filter is not sufficiently large \cite{Bqg:07} and this phenomenon gradually saturates the performance gain due to FS \cite{Tcc:04,Nts:07}. Therefore, it is important to determine the operating point for the FS rate, a design parameter that we can actually configure to find a good trade-off between performance and complexity. This is the third contribution of this paper.

The remainder of this paper is structured as follows: Section II reviews in detail prior work in the areas of eigenvalue based sensing. Section III describes the considered signal models under white or correlated noise scenarios. Section IV analyzes the effect of noise correlation for the noise only case and proposes new SCN-based decision bounds. Section V provides the analysis for signal plus noise case under white or correlated noise and describes the purposed eigenvalue based SNR estimation method. Section VI studies the performance of proposed techniques with numerical simulations and proposes a method
for determining the optimal FS operating point. Section VII concludes the paper. The appendix includes some preliminaries on random matrix transforms.
\subsection{Notation}
Throughout the formulations of this paper, \(\mathbb{E}[\cdot]\) denotes the expectation, \(\left(\cdot\right)^\dagger\) denotes the conjugate transpose matrix, $\mathrm{abs}(\cdot)$ denotes the absolute value, and $(\cdot)^H$ denotes the Hermitian transpose matrix, $\mathcal S_{\mathbf X}$ represents Stieltjes transform of $ \mathbf X$, $\mathcal R_{\mathbf X}$ represents R transform and $\Sigma_{\mathbf X}$ represents $\Sigma$ transform \cite{Ant:04}.
\section{Related Work}
The three major eigenvalue based sensing techniques considered in the literature are \cite{Yyl:09}: Maximum-Minimum Eigenvalue (MME) detection, Energy with Minimum Eigenvalue (EME) detection and Maximum Eigenvalue Detection (MED). A number of eigenvalue based SS methods are proposed in \cite{Yyl:09,Krs:11,Cdb:08} utilizing eigenvalue properties of Wishart random matrices, which arise under noise only cases in white noise scenarios. The authors in \cite{Cdb:08} use MP law to test binary hypothesis problem. In \cite{Yyl:09}, the Tracy-Widom (TW) distribution is used as a statistical model for the largest eigenvalue and both the TW distribution and the MP models are used to find the approximate distribution of random SCN and this distribution is used to derive the relationship between an expression for probability of false alarm ($P_f$) and threshold. The difference between the MP approach and the TW approach is that MP is a deterministic function which characterizes the asymptotic matrix spectrum, while TW approach provides the statistics of individual eigenvalues e.g. the maximum eigenvalue. Since the rate of convergence of the TW distribution is faster than MP law, the TW method is superior than the MP only method. However, the TW method outperforms the MP method only at relatively large SNRs since SCN is a ratio of two random variables and the approximation considered in \cite{Yyl:09} is accurate only for large SNR conditions.

In \cite{Krs:11}, an approximation of the threshold function is derived for systems having equal number of receiving antennas and samples. In \cite{Pgf:09}, the p.d.f. of the eigenvalue ratio has been derived using the expression of the joint distributions of an arbitrary subset of ordered eigenvalues of complex Wishart matrices. In this scenario, the receiver should be provided with a look-up table in order to calculate the proposed inverse cumulative distribution function of the second-order TW distribution. The exact distribution of the condition number of a complex Wishart matrix has been used to calculate the threshold expression in \cite{Krs:11} without the need of a look up table. However, the calculated threshold expression in terms of $P_f$ in \cite{Krs:11} is based on the exact density of the condition number of complex Wishart matrix considering the noise only case and it is only valid in case of white noise measurements. For the correlated noise case, the sample covariance matrix does not follow the properties of Wishart random matrices.

In \cite{Fga:09}, a more accurate model considering the Trace-Widom-Curtiss (TWC) model has been considered by using the distribution of the smallest eigenvalues of Wishart random matrices. However, the TW distribution and the Curtiss' ratio of variates formula are highly involved functions, which are hard to evaluate numerically and non tractable to find the support of a.e.p.d.f. \cite{Wsg:11}. In \cite{Wsg:11}, the exact distribution of SCNs of dual Wishart random matrices has been used and it is argued that the proposed method requires only tens of samples and outperforms all the RMT based techniques. However, the authors in \cite{Wsg:11} considered the Wishart random matrix model for signal plus noise case for simplicity and did not address the fact that during the presence of signal and correlated noise, the sample covariance matrix may no longer be a Wishart random matrix. In \cite{Lbj:11}, non-asymptotic behavior of eigenvalues of random matrices has been considered using the spectral properties of random sub-Gaussian matrices of fixed dimensions. A cooperative SS algorithm using double eigenvalue threshold has been proposed in \cite{Kzh:10}, which considers two maximum eigenvalues for the noise only and the signal plus noise cases through analysis of sample covariance matrix of received signals using RMT approach.

Spectrum sensing using free probability theory has also received important attention in the literature \cite{Lbj:09} \cite{Lbj:10}. In \cite{Lbj:09}, a cooperative scheme for SS has been proposed using asymptotic free behavior of random matrices and the property of Wishart distribution. The same work has been extended for MIMO scenario in \cite{Lbj:10}. In these works, the presence of the PU signal is decided simply by checking whether the power matrix is zero or not but this technique is not studied analytically in \cite{Lbj:09} and \cite{Lbj:10}.
\section{Signal Model}
Let us consider a single cognitive user and a single primary user for simplicity of analysis. Let $N$ be the number of samples analyzed by the cognitive user for the decision process in the time duration of $\tau$ while performing symbol rate sampling. The sampling rate in the receiver can be increased beyond the symbol rate to enhance the signal detection capability in wireless fading channels. This technique known as FS \cite{Tcc:04} produces $N$ FS samples out of each original sample. Let $M$ be the FS rate carried out at the input of cognitive receiver. From signal model point of view, this factor can be considered as the number of multiple outputs analogous to the number of cooperating users in cooperative based sensing or the number of antennas in multiple antenna sensing as considered in related literature \cite{Yyl:09,Cdb:08}. We denote the hypotheses of the presence and absence of the PU signal by $H_1$ and $H_0$ respectively. A binary hypothesis testing problem for $k$-th FS branch, $k=1,...,M$, can be written as:
 \begin{eqnarray}
  H_0: y_k(i) &=& \hat{z}_k(i) \hspace{65 pt} \mathrm{PU \hspace{5 pt}  absent} \nonumber \\
  H_1: y_k(i) &=& h_k(i) s(i)+\hat{z}_k(i) \hspace{10 pt} \mathrm{PU \hspace{5 pt}  present}
 \end{eqnarray} where $y_k(i)$ is the signal observed by a $k$-th receiving dimension at $i$-th instant, $i=1,2,..,N$, $s(i)$ is the PU signal at $i$-th instant, which is to be detected, $h_k(i)$ is the amplitude gain of the channel for $k$-th receive dimension at $i$-th instant, and $\hat{z}_k(i)$ denotes the colored noise for $k$-th receive dimension at $i$-th instant. For our analysis, we assume that
transmitted symbols are independent and identically distributed (i.i.d.) complex circularly symmetric (c.c.s.) Gaussian symbols, the noise samples in each FS branch are independent but are correlated across
FS branches.

The  $M \times N$ channel matrix $\mathbf H$ consists of i.i.d. coefficients and each row of $\mathbf H$ represents the channel coefficients for $N$ number of samples for each FS branch i.e. $\textbf{H} \triangleq [\textbf{h}_1^T, \textbf{h}_2^T,...,\textbf{h}_M^T]^T$, with $\textbf{h}_m \triangleq [\begin{array}{cccc}
             h_m(1) & h_m(2) & \ldots & h_m(N)
\end{array}]$ with $m=1,2,...,M$. We assume channel coefficients to be i.i.d. in each FS branch and the channel coherence time to be sufficiently small so that channel is not
correlated as we increase the FS rate.

While performing sensing in a cognitive receiver, the sensing duration ($\tau$) and symbol interval ($T_s$) may not be the same depending upon the signal bandwidth and sampling rate used at the receiver. For example, let us consider a coexistence scenario of TV whitespace broadband and wireless microphone systems. These are two systems with different operation bandwidths, a microphone signal typically occupies 200 kHz bandwidth while TV signal occupies 6 MHz and microphone operates on TV bands \cite{Yyl:09}. In this scenario, $\tau$ becomes much greater than $T_s$. Under the $H_1$ hypothesis, we consider the following signal models considering the relation between $\tau$ and $T_s$.\\
\textit{Case 1}: In this case, we consider that the transmitted symbol remains constant during the sensing period. The received signal matrix in this case can be written as: $\textbf{Y}= \sqrt{p} \textbf{H} s+\hat{\mathbf{Z}}$, where $s$ is a constant transmitted symbol, $p$ is the power of transmitted symbol and $\hat{\mathbf{Z}} \triangleq [\hat{\mathbf{z}}_1^T, \hat{\mathbf{z}}_2^T,...,\hat{\mathbf{z}}_M^T]^T$,
with $\hat{\mathbf{z}}_m \triangleq [\begin{array}{cccc}
             \hat{z}_m(1) & \hat{z}_m(2) & \ldots & \hat{z}_m(N)
\end{array}]$ . In this case, the sample covariance of transmitted signal can be written as: $R_s=\mathbb E[s^2]=1$.\\
\textit{Case 2}: In this case, each fractional sampled branch i.e. each row of matrix $\mathbf Y$ includes the samples for a single symbol. $\textbf{Y} = \sqrt{p} \textbf{H} \textbf{S}_d+\hat{\mathbf{Z}}$, where $\textbf{S}_d$ is the diagonal transmitted signal matrix of dimension $N \times N$ with diagonal $\mathbf{s}=[s(1)...s(N)]$. In this case, the sample covariance matrix of the transmitted signal becomes
\begin{equation}\label{}
  \textbf{R}_{\mathbf S}=\mathbb{E}[\textbf{S}_d\textbf{S}_d^H]=
 \left[
  \begin{array}{ccccc}
    \mathbb{E}[s^2(1)] & 0 & \cdots  & 0 \\
    0 & \mathbb{E}[s^2(2)] & \cdots  & 0 \\
      &  &  \ddots  &  \\
     0 & 0 & \cdots & \mathbb{E}[s^2(N)] \\
  \end{array}
\right]
= \mathbf{I}
\end{equation} assuming that for each sample we get an i.i.d. c.c.s. Gaussian symbol with $\mathbb E[s^2]=1$. \\
The $M \times N$ received signal matrix $\textbf{Y}$ in both cases can be written in the following form:
 \begin{equation}\label{}
    \textbf{Y}=\left[\begin{array}{c}
                 \textbf{y}_1 \\
                 \textbf{y}_2 \\
                 \vdots \\
                 \textbf{y}_M
               \end{array}\right]=\left[
                                    \begin{array}{cccc}
                                      y_1(1) & y_1(2) & \ldots & y_1(N) \\
                                      y_2(1) & y_2(2) & \ldots & y_2(N) \\
                                      \vdots & \vdots & \ddots & \vdots \\
                                      y_M(1) & y_M(2) & \ldots & y_M(N)\\
                                    \end{array}
                                  \right]
 \end{equation}

Assuming that the source signal is independent from the noise, the covariance matrix of received signal, $\textbf{R}_{\mathbf Y}$, is given by \cite{Yyl:09};
\begin{equation}\label{}
   \textbf{R}_{\mathbf Y}= \mathbb{E}[\textbf{YY}^H] =\mathbb{E}\left[(\sqrt{p} \textbf{HS})(\sqrt{p} \textbf{HS})^H \right]+ \mathbb{E}[\hat{\mathbf{Z}} \hat{\mathbf{Z}}^H]=p \textbf{H} \textbf{H}^H + \textbf{R}_{\hat{\mathbf Z}}
\end{equation} where $\textbf{R}_{\hat{\mathbf Z}}=\mathbb{E} [\hat{\mathbf{Z}}\hat{\mathbf{Z}}^H]$. Let us define sample covariance matrices of received signal and noise as: $\hat{\textbf{R}}_{\mathbf Y}(N)= \frac{1}{N} \textbf{Y}\textbf{Y}^H$ and $\hat{\textbf{R}}_{\hat{\mathbf Z}}(N)=\frac{1}{N} \hat{\mathbf{Z}}\hat{\mathbf{Z}}^H$.
The received signal $\mathbf{Y}$ can be further written as:
\begin{equation}\label{}
    \textbf{Y} = \left\{
             \begin{array}{ll}
                \sqrt{p} \textbf{H}s + \hat{\textbf{Z}}, & \mathrm{Case} \hspace{5 pt} {1}\\
                \sqrt{p} \textbf{H} \textbf{S}_d + \hat{\textbf{Z}}, & \mathrm{Case} \hspace{5 pt} {2}\\
             \end{array}
             \right.
             \label{eq: rxs}
\end{equation}
where $\hat{\textbf{Z}} \sim \mathcal{CN}(0, \hat{\textbf{R}}_{\hat{\mathbf Z}}(N))$ is the colored noise. The SCN of $\hat{\textbf{R}}_{\hat{\mathbf Z}}(N)$ depends on the noise correlation among noise samples
across FS branches.
\vspace{-5 pt}
\subsection{Noise Correlation Modeling}
To analyze the noise correlation effect mathematically, a simple correlation model should be employed.
In this work, we consider one-sided noise correlation model. Using this model, the colored noise can be related to the white noise using the following expression.
\begin{equation}\label{}
  \hat{\textbf{Z}}=\mathbf{\Theta}^{1/2}\textbf{Z}
\end{equation} where $\textbf{Z}$ is an $M \times N$ matrix
with c.c.s. i.i.d. Gaussian entries with zero mean and unit variance, representing the white noise and $\mathbf{\Theta}$ is an $M \times M$ Hermitian matrix whose entries correspond to the correlation
among noise samples across FS branches. To ensure that $\mathbf \Theta$ does not affect the noise power, we consider the following normalization:
\begin{equation}\label{}
  (1/M) \mathrm{trace}\{\mathbf {\mathbf \Theta}\}=1
\end{equation}
The noise covariance matrix depends on the transfer function of the pulse shaping filter used at the input of RF front end of a CR. Most SS-related papers consider baseband processing without including the effect of the filter on the received signal and they further assume the noise to be white in many scenarios. In practical implementation of a CR, the received signal should be passed through a pulse shaping filter before further processing and thus noise is no longer white. Due to absence of channel knowledge and PU signal, coherent receivers such as matched filter (i.e. receive part of root raised cosine filter) are not suitable for the SS application. Active RC filters with tunable cut off frequencies has been proposed in literature for CR applications \cite{Vvt:06} \cite{Hyk:10}. When a white noise input process with power spectral density $N_0/2$ is the input to a RC filter with time constant $\mathrm{RC}$, the noise is colored after filtering. Although the channel may also get correlated at the
output of the filter, we are interested in analyzing the effect of noise correlation on SS performance in this work assuming that noise correlation effect dominates the overall effect. The RC filter transforms the input autocorrelation function of white noise into an exponential function given by \cite{Shaykin}: $R_y(\nu)=\frac{N_0}{4 RC}e^{-\frac{|\nu|}{RC}}$. Since the autocorrelation function of output process of RC filter resembles the exponential model, we consider exponential correlation to model $\mathbf{\Theta}$ in this work. The exponential correlation model can be written as \cite{Sll:01,Csh:09}:
\begin{equation}\label{}
    \theta_{ij} \sim \left\{
             \begin{array}{ll}
               \rho^{\mathrm{abs}(j-i)}, & i \leq j \\
               \left({\rho}^{\mathrm{abs}(j-i)}\right)^\dagger, & i > j
             \end{array}
           \right.
\end{equation} where $\theta_{ij}$ is the ($i,j$)-th element of $\mathbf \Theta$ and $\rho \in \mathcal{C}$ is the correlation coefficient with $\mid \rho \mid \leq 1$.
\section{Analysis under $H_0$ hypothesis }
RMT has been used in the literature in various applications such as modeling transmit/receive correlation in MIMO channels and multiuser MIMO fading \cite{Csh:09,Mfp:03}. Here, we state two RMT based theorems which are going to be used in defining our decision statistics.
 \begin{thm}\cite{Ant:04} Consider an $M \times N$ matrix $\textbf{F}$ whose entries are independent zero-mean complex (or real) random variables with variance $\frac{1}{N}$ and fourth moments of order $O\left(\frac{1}{N^2}\right)$. As $M,N \rightarrow \infty$ with $\frac{N}{M} \rightarrow \beta$, the empirical distribution of the eigenvalues of $\frac{1}{N}\textbf{F}\textbf{F}^H$ converges almost surely to a non random limiting distribution with density given by:
\begin{equation}\label{}
    f_\beta(\lambda)= \left( 1 - \frac{1}{\beta} \right)^{+} \delta(\lambda) + \frac{ \sqrt{(\lambda-a)^{+} (b-\lambda)^{+}}} {2 \pi \beta \lambda}
\end{equation} where $a=(1-\sqrt{\beta})^2$, $b=(1+\sqrt{\beta})^2$. The parameters $a$ and $b$ define the support of the distribution and correspond to the minimum eigenvalue ($\lambda_{min}$) and the maximum eigenvalue ($\lambda_{max}$) respectively and the ratio $b/a$ defines the SCN of $\frac{1}{N}\textbf{F}\textbf{F}^H$. The above limiting distribution is the MP law with ratio index $\beta$.
\end{thm} \qed
\begin{rmk}
\textit{In practice, we can have only finite number of samples and the sample covariance matrix $\hat{\textbf{R}}_{\mathbf Y}(N)$ may deviate from the covariance matrix $\textbf{R}_{\mathbf Y}$ \cite{Yyl:09}. The eigenvalue distribution of $\hat{\textbf{R}}_{\mathbf Y}(N)$ becomes complicated due to requirement of consideration of finite parameters in the analysis. This makes the choice of the threshold difficult for SS purpose and the performance of SS algorithms becomes sensitive to the choice of threshold at low values of SNR. Although various TW approaches have been proposed in \cite{Yyl:09} and \cite{Fga:09} for accounting the random nature of SCN of finite matrices, we are interested in analyzing the correlation effect on MP based asymptotic methods in this paper.}
\end{rmk}
In this noise only case, $\hat{\textbf{R}}_{\mathbf Y}(N)$ becomes equal to $\hat{\textbf{R}}_{\hat{\mathbf Z}}(N)$ and can be written as:
\begin{equation}\label{}
    \hat{\textbf{R}}_{\mathbf Y}(N)=  \hat{\textbf{R}}_{\hat{\mathbf Z}}(N)=\mathbf{\Theta}^{1/2} \textbf{ZZ}^H \mathbf{\Theta}^{1/2}
\end{equation} $\hat{\textbf{R}}_{\mathbf Y}(N)$ converges to $\textbf{R}_{\mathbf Y}$ for $N \rightarrow \infty$ \cite{Fga:09} and asymptotic analysis still holds true for large values of $N$ \cite{Cdb:08}. Furthermore, $\hat{\textbf{R}}_{\mathbf Z}(N)=\frac{1}{N} \textbf{Z}\textbf{Z}^H$ is nearly a Wishart random matrix \cite{Ant:04} in white noise scenarios but is no longer a Wishart random matrix in correlated noise scenarios.

To calculate the threshold for SS purpose, we need the support of a.e.p.d.f. of \textbf{Y}, namely, $\lambda_{max}$ and $\lambda_{min}$. Due to noncommutative nature of random matrices, it is not straightforward to calculate the eigenvalue distribution of $\textbf{Y}$ by knowing the eigenvalue distribution of $\mathbf{\Theta}$ and $\textbf{Z}$. Using free probability analysis, the asymptotic eigenvalue distribution of $\frac{1}{N}\textbf{Y}\textbf{Y}^H$ can be calculated by applying the $\Sigma$ transform \cite{Ant:04}
  \begin{equation}\label{}
   \Sigma_{\mathbf{Y}}(z)= \Sigma_\mathbf{\Theta}(z) \Sigma_{\textbf{Z}}(z)
\end{equation} where $\Sigma_\mathbf{\Theta}$ and $\Sigma_{\textbf{Z}}$ are the $\mathbf \Sigma$ transforms of the densities of eigenvalues of $\mathbf{\Theta}$ and $\frac{1}{N}\textbf{ZZ}^H$ respectively. Since $\mathbf{\Theta}$ is a square matrix, $\mathbf{\Theta}^{1/2} \textbf{Z}\textbf{Z}^H \mathbf{\Theta}^{1/2}$ and $\mathbf{\Theta} \textbf{Z}\textbf{Z}^H$ have identical eigenvalues \cite{Ant:04}.
Using eqn. (\ref{eq: Str}), $\Sigma_{\mathbf Y}(z)$ can be written as:
 \begin{equation}\label{}
   \Sigma_{\mathbf{Y}}(z)= \Sigma_\mathbf{\Theta}(z) \frac{1}{z+\beta}
\end{equation}
The $\Sigma$ transform of corresponding asymptotic eigenvalue distribution ${\Sigma}_\mathbf{\Theta}(z)$ can be obtained by choosing a proper model for noise correlation. The asymptotic density of eigenvalues of $\mathbf \Theta$ can be described as a tilted semicircular law \cite{Mfp:03}, which is a close approximation for the exponential model and is analytically tractable. This density can be described using the following theorem.
\begin{thm}
\cite{Mfp:03} Let $\mathbf{\Theta}$ be a positive definite matrix which is normalized as: $(1/M) \mathrm{trace}\{\mathbf {\mathbf \Theta}\}=1$, and whose asymptotic spectrum has the p.d.f.
\begin{equation}\label{}
    f_{\mathbf{\Theta}} (\lambda)= \frac{1}{2 \pi \mu \lambda^2} \sqrt{\left(\frac{\lambda}{\sigma_1}-1 \right) \left(1-\frac{\lambda}{\sigma_2}\right)}
\end{equation}  with $\sigma_1 \leq \lambda \leq \sigma_2 $ and $\mu = \frac{(\sqrt{\sigma_2}-{\sqrt{\sigma_1}})^2}{4 \sigma_1 \sigma_2}$. If $\textbf{F}$ is an $M \times N$ standard Gaussian matrix as defined in theorem 1, then as $M,N \rightarrow \infty$ with $\frac{N}{M} \rightarrow \beta$, the asymptotic distribution of $\mathbf W= \mathbf{\Theta}^{1/2} {\textbf{FF}}^H \mathbf{\Theta}^{1/2}$ has the following p.d.f.
\begin{equation}\label{}
    f_{\textbf{W}} (\lambda) = (1-\beta)^+ \delta(\lambda) + \frac{\sqrt{(\lambda-\tilde{a})^+(\tilde{b}-\lambda)^+}}{2 \pi \lambda (1+\lambda \mu)}
\end{equation} where
\begin{eqnarray}
  \tilde{a} &=& 1+\beta+2 \mu \beta-2 \sqrt{\beta} \sqrt{(1+\mu)(1+\mu \beta)}  \nonumber \\
  \tilde{b} &=& 1+\beta+2 \mu \beta + 2 \sqrt{\beta} \sqrt{(1+\mu)(1+\mu \beta)}
\end{eqnarray} The parameters $\tilde{a}$ and $\tilde{b}$ correspond to $\lambda_{min}$ and $\lambda_{max}$ respectively and the ratio $\tilde{b}/\tilde{a}$ defines the SCN of $\mathbf W$.
\end{thm} \qed \\
The eigenvalue spread of $\mathbf{\Theta}$ is related to the degree of noise correlation i.e. a zero eigenvalue spread corresponds to a zero-correlation model $\mathbf{\Theta}=\textbf{I}_M$ and higher spreads are associated with higher correlation modes. In (14), the parameter $\mu$ controls the degree of noise correlation and varies the support of the distribution i.e. for $\mu=0$, $\tilde{a}=a$ and $\tilde{b}=b$. For the exponential correlation model as stated in \cite{Mfp:03}, the parameter $\mu$ is related to correlation coefficient $\rho$ with the following relation:
$\mu=\frac{\rho^2}{1-\rho^2}$. Furthermore, the SCN is related to $\rho$ with the relation $\mathrm {SCN}=\frac{1+\rho}{1-\rho}$.
To calculate $\mu$ in a practical cognitive receiver, the value of $\rho$ can be determined from FS rate based on some empirical model constructed from measurements. In our results, we employ a simple linear model to study the effect of noise correlation as FS rate increases (see Section VI C).

It can be noted that MP law can be used as a theoretical prediction under the $H_0$ hypothesis with white noise \cite{Cdb:08}. The support of the eigenvalues of the sample covariance matrix under the $H_0$ hypothesis is finite independently of the distribution of the noise. To decide the absence or presence of signal under white noise, the deviations of distribution of eigenvalues from the normal bounds $a$ and $b$ of MP law can be used. If the eigenvalues appear outside these bounds, then it can be decided that there is presence of PU signal and if all the eigenvalues lie within the bounds of MP law, it can be decided that there is absence of PU signal. In case of noise correlation, the bounds of eigenvalue distribution of sample covariance matrix become different than the bounds obtained in white noise case and MP law no longer applies. The new bounds $(\tilde{a},\tilde{b})$ depend on the noise correlation parameter $\mu$. We present the sensing example with new bounds for FS scenario in the following subsection.
\vspace{-5 pt}
\subsection{Sensing With FS}
The parameter $\mu$ depends on the sampling rate applied in the receiver since noise correlation increases along with the sampling rate. Sampling rate can be varied from the symbol rate to some order of the symbol rate and the effect of sampling rate on sensing performance can be evaluated by varying the correlation effect. Let us consider that both noise distribution and noise variance are unknown to the detector to reflect the practical scenario. The SCN under the $H_0$ hypothesis does not depend on noise variance. Under white noise scenario, $H_0$ hypothesis is decided if $\mathrm {SCN} \leq \frac{b}{a}$ and $H_1$ is decided in all other conditions \cite{Cdb:08}. Similar decision process can be applied for correlated noise case and the decision can be made on the basis of the following condition:
\begin{equation}\label{}
    \mathrm{decision} = \left\{
             \begin{array}{ll}
                H_0 , &  \mathrm{if} \hspace{5 pt} \mathrm {SCN} \leq \frac{{\tilde{b}}}{{\tilde{a}}}\\
                H_1 , & \mathrm{Otherwise}
             \end{array}
             \right.
             \label {eq: decs}
\end{equation}
When FS rate $M$ is applied at the CR, $M$ rows of sample covariance matrix become correlated. The value $M=1$ corresponds to the symbol rate sampling and correlation coefficient $\rho=0$. Since the the value of $\rho$ varies from $0$ to $1$, the relation between the FS rate $M$ and the correlation coefficient $\rho$ is considered as a simple linear model \footnote{More exact relation models can be acquired through measurements
on the CR equipment.} as shown below
\begin{equation}\label{}
   \rho = \varepsilon \left(\frac{1}{\beta}-\frac{1}{N} \right)
   \label{eq: linear}
\end{equation} where $\beta=N/M$ and $\varepsilon$ is a parameter defining the slope of the linear dependence.

\section{Analysis under $H_1$ hypothesis}

\subsection{White Noise}
Assuming that signal and noise are independent, for very large value of $N$, eqn. (\ref{eq: rxs}) leads to the following approximation \cite{Yyl:09}.
\begin{equation}\label{}
   \hat{\textbf{R}}_{\mathbf Y}(N)\approx  p \textbf{H} \textbf{H}^H + \hat{\textbf{R}}_{\mathbf Z} (N)
\end{equation}
In white noise case, the sample covariance of received signal under assumed conditions can be realized as the sum of two Wishart matrices i.e. $p \textbf{H} \textbf{H}^H$ and $\textbf{Z} \textbf{Z}^H$ with same degree of freedom and different covariance structures. In this condition, MP law holds true for both matrices. Although it is possible to find another Wishart matrix from the the addition of $p \textbf{H} \textbf{H}^H$ and $\textbf{Z} \textbf{Z}^H$ approximately (see Lemma 6, \cite{Nah:11}) and then apply scaled MP law by scaling with variance $(1+p^2)$ for the new Wishart matrix, we use free probability theory for more accurate analysis. The R transform of eigenvalue density function of $\textbf{Y} \textbf{Y}^H$ can be found by adding the R transforms of density functions of $p \textbf{H} \textbf{H}^H$ and $\textbf{Z} \textbf{Z}^H$ using free probability theory.
Using eqn. (\ref{eq: Rmpp}), the R transform of $p\textbf{H}\textbf{H}^H$ can be written as:
\begin{equation}\label{}
   \mathcal{R}_{p\textbf{H}\textbf{H}^H}(z)=p \mathcal{R}_{\textbf{H}\textbf{H}^H}(pz)= \frac{p \beta }{1-p z}
\end{equation} Since the R transform of $\textbf{Z} \textbf{Z}^H$ is $\mathcal{R}_{\textbf{Z} \textbf{Z}^H}(z)=\frac{\beta}{1- z}$ from eqn. (\ref{eq: Rmp}), the combined R transform can be written as:
\begin{equation}\label{}
   \mathcal{R}_{{\textbf{Y}\textbf{Y}^H}}(z)= \frac{p \beta}{1-p z}+\frac{\beta}{1- z}
   \label{eq: Rtrans}
\end{equation}
The inverse Stieltjes transform can be obtained by applying eqn. (\ref{eq: Rtrans}) on eqn. (\ref{eq: Sttransform}). Then the Stieltjes transform can be obtained by solving the following cubic polymonial.
\begin{eqnarray}
(y p) \mathcal{S}(z) ^ 3 + ( p(-2 \beta+z+1)+z) \mathcal{S}(z) ^ 2 + ( (1-\beta)(1-p)+z) \mathcal{S}(z)+1
\label{eq: uncor}
\end{eqnarray}
\subsection{Correlated Noise}
Using the similar arguments as in the above subsection, the following approximation can be written for correlated noise scenario.
\begin{equation}\label{}
   \hat{\textbf{R}}_{\mathbf Y}(N)\approx  p \textbf{H}\textbf{H}^H + \hat{\textbf{R}}_{\hat{\mathbf Z}}(N)
\end{equation}
In correlated noise case, the sample covariance of received signal under assumed conditions can be realized as a sum of one Wishart matrix  i.e. $p \textbf{H} \textbf{H}^H$ and another correlated Wishart matrix $\hat{\textbf{Z} } \hat{\textbf{Z}}^H$. In this condition, MP law can be applied for $p \textbf{H} \textbf{H}^H$ and the analysis carried out under $H_0$ hypothesis in section IV can be applied for $\hat{\textbf{Z} } \hat{\textbf{Z}}^H$. Then the R transform of density function of the received signal can be found by adding the R transforms of density functions of $p \textbf{H} \textbf{H}^H$ and $\hat{\textbf{Z} } \hat{\textbf{Z}}^H$. The Stieltjes transform of $\hat{\textbf{Z} } \hat{\textbf{Z}}^H$ can be written as \cite{Mfp:03}:
\begin{equation}\label{}
\mathcal S_{\hat{\textbf{Z} } \hat{\textbf{Z}}^H}(z)=\frac {z+2z \mu+1-\beta+\sqrt{[z-(1+\beta)]^2-4 \beta(1+\mu z)}} {2z(1+\mu z)}
\end{equation}
Then the R transform for $\hat{\textbf{Z} } \hat{\textbf{Z}}^H$  is calculated using eqn. (\ref{eq: Sttransform}) and can be expressed as:
\begin{equation}\label{}
  \mathcal {R}_{\hat{\textbf{Z} } \hat{\textbf{Z}}^H}(z)=-\frac{1}{2}\frac{(z-1+\sqrt{(z^2-2 z+1-4 \mu \beta z)})}{\mu z}
\end{equation}
The combined R transform then becomes
\begin{equation}\label{}
\mathcal{R}_{\mathbf{Y} \mathbf{Y}^H}(z)= \frac{p \beta}{(1- p z)} - \frac{1}{2} \frac{(-1+z+\sqrt{(1-2z+z^2-4 \mu \beta z)})}{z \mu}
\label{eq: Rtrans1}
\end{equation}
The inverse Stieltjes transform can be obtained by applying eqn. (\ref{eq: Rtrans1}) on eqn. (\ref{eq: Sttransform}) and the Stieltjes transform can be obtained by solving the following quatric polymonial.
{\begin{eqnarray}
 (z p ^2(1+\mu z)) \mathcal{S}(z) ^ 4 + (2 z \mu p (z-p \beta) + p ^ 2(1+2z \mu +z-2 \beta) +2 z p)  \mathcal{S}(z)^ 3 + \nonumber \\
  (p ^ 2 (\mu (1-\beta)^2+1-\beta)+2p(1+z+\mu z (2-\beta))+z-3p \beta+z^2 \mu )\mathcal{S}(z)^2+ \nonumber \\
   (2 p(1+\mu(1-\beta))+z(1+2 \mu)-\beta (1+p)+1) \mathcal{S}(z)  + 1 + \mu
 \label{eq: corrp}
\end{eqnarray}}
\normalsize{\baselineskip24pt
We can find the roots of the above polymonial in closed form. The closed form is not specifically written in this paper because it includes many terms which provide no further insight. The support of a.e.p.d.f. of $\frac{1}{N}\mathbf{YY}^H$ under correlated noise is calculated based on eqn. (\ref{eq: corrp}). Since we know the value of $\beta$ and we can measure the value of $\rho$, we can find the value of $p$ by sensing the maximum value of $\frac{1}{N}\mathbf{YY}^H$.
Look up tables are provided for convenience in order to estimate the SNR of the PU (see Section VI B). We consider the following three cases: (i) signal plus correlated noise, (ii) correlated noise only, and (iii) signal plus white noise. In the look up table, we present the maximum eigenvalues of received signal's covariance matrix for above three cases for different values of SNR and $\beta$. The eigenvalue distribution for signal plus white noise case is obtained using polymonial eqn. ($\ref{eq: uncor}$) and for signal plus correlated noise case is obtained using polymonial eqn. ($\ref{eq: corrp}$). We can predict the received SNR of PU signal based on the maximum eigenvalue. The parameters $\beta$ and $\rho$ are assumed known as operating parameters of the sensing module. Based on this
estimated SNR, we could potentially design suitable underlay transmission strategy for secondary transmission. In section VI, we provide the normalized  MSE versus SNR plot (see Fig. 7) to evaluate
the performance of this estimation technique.
\section{Numerical Results}
In this section, we study the performance of eigenvalue based sensing in the presence of noise correlation with proposed decision bounds. The decision statistic for MP law is calculated as:
 \begin{equation}\label{}
    \mathrm{decision} = \left\{
             \begin{array}{ll}
                H_0 , &  \mathrm{if} \hspace{5 pt} \mathrm {SCN} \leq \frac{{b}}{{a}}\\
                H_1 , & \mathrm{Otherwise}
             \end{array}
             \right.
             \label{eq: decmp}
\end{equation} and proposed decision statistic is calculated based on eqn. (\ref{eq: decs}), which can be rewritten as:
\begin{equation}\label{}
    \mathrm{decision} = \left\{
             \begin{array}{ll}
                H_0 , &  \mathrm{if} \hspace{5 pt} \mathrm {SCN} \leq \frac{\tilde{b}}{\tilde{a}}\\
                H_1 , & \mathrm{Otherwise}
             \end{array}
             \right.
             \label{eq: decmp1}
\end{equation}

The ratio of correct sensing is used as a performance metric to analyze the performance and it is defined as the ratio of number of correct sensing to the number of total considered iterations under both hypotheses. In the presented simulation results, $10^3$ iterations were considered.

Furthermore, we present SNR estimation method under $H_1$ hypothesis. The normalized MSE is considered as a parameter to characterize the performance of proposed SNR estimation technique and is defined as:
\begin{equation}\label{}
 \mathrm {MSE}=\frac{\mathbb E[\hat{\gamma}-\gamma]^2}{\gamma^2}
\end{equation} where $\hat{\gamma}$ is the estimated SNR with the proposed method and $\gamma$ is the actual SNR.

\subsection{Eigenvalue SS}
The performance of proposed sensing scheme has been analyzed in Rayleigh fading channel. White and correlated noise scenarios have been considered in the analysis. In case of white noise scenarios, it has been noted that the eigenvalue distribution of received signal's covariance matrix follows MP law and the distribution is limited to the bounds given by MP law. Therefore, the decision rule in eqn. (\ref{eq: decmp}) is used for sensing of the PU signal. However, in case of noise correlation, the eigenvalue distribution breaks into two parts (Fig. 2, \cite{Csh:09}) and the decision rule in eqn. (\ref{eq: decmp1})
is considered. The ratio of correct sensing versus SNR for $\rho=0.5, \beta=1/6$ is depicted in Fig. 1. It can be observed that sensing with eqn. (\ref{eq: decmp1}) outperforms than sensing with eqn. (\ref{eq: decmp}) in correlated noise case. Figure 2 shows the sensing performance versus correlation coefficient at SNR value of -6 dB and $\beta=1/6$ and it can be noted that with increased amount of noise correlation, the sensing with MP bounds decreases drastically and sensing with eqn. (\ref{eq: decmp1}) gives better performance up to some value of correlation. Moreover, it has been noted that new bounds also do not provide better sensing at high correlation region. This is due to the fact that the threshold increases and the asymptotic eigenvalue support of $H_1$ is subsumed in the one of $H_0$ at this region.
To study the effect of correlation on eigenvalue based spectrum sensing under $H_1$ hypothesis, simulations were carried out in correlated and white noise scenarios. $H_1$ hypothesis case was considered by taking the combination of signal and noise in both scenarios. Figure 3 shows the theoretical and simulated eigenvalue distribution of covariance matrix of received signal i.e. $\frac{1}{N}\textbf{Y}\textbf{Y}^H$ for SNR=-2 dB and $\beta$=1 under white noise case. The histograms of the eigenvalues were created by accumulating the eigenvalues over $10^3$ iterations. The theoretical result was obtained by evaluating the polynomial given by eqn. (\ref {eq: uncor}). Similarly, Fig. 4 shows the eigenvalue distribution of covariance matrix of received signal for SNR=-2 dB, SCN=3 and $\beta$=1 under correlated noise. In this case, theoretical result was obtained by evaluating the polynomial given by eqn. (\ref{eq: corrp}). From figures 3 and 4, it can be observed that the theoretical and simulated density functions perfectly match.

To observe the variation of the maximum eigenvalue of received signal's covariance matrix with respect to SNR, we present the maximum eigenvalue versus SNR plot in Fig. 5 for both correlated and white noise cases. From the figure, it can be observed that the maximum eigenvalue has higher value in correlated scenario than in white noise scenario over the considered range of SNR (from -10 dB to 2 dB) and the gap between these two curves goes on decreasing while increasing the value of SNR.

Figure 6 shows the plot of maximum eigenvalue of received signal's covariance matrix versus SCN of correlation matrix for the following three cases (i) signal plus correlated noise case, (ii) correlated noise only case and (iii) signal plus white noise case. It can be observed that the maximum eigenvalue for the first case is greater than maximum eigenvalue for the second case and the difference remains more or less consistent for all considered values of SCN (from 2 to 20). With respect to white noise case, maximum eigenvalue in correlated case increases almost linearly with the value of SCN.

\subsection{SNR Estimation}
Table 1 shows the look up table for different values of SCNs of the correlation matrix. This table can be used to estimate the SNR of the PU signal based on the values of SCN and $\beta$ for both correlated and white noise cases. The value of SCN can be derived from the measurements of $\rho$. For example, if the value of SCN is 2, $\beta$ is 1 and maximum eigenvalue of covariance matrix of received signal i.e. $\frac{1}{N}\textbf{Y}\textbf{Y}^H$ is $5.93$, we can estimate that the received SNR of PU signal is 0 dB and intermediate values can be calculated through interpolation. Based on this estimated SNR, we can
design suitable underlay transmission strategy for secondary transmission. From the table, it can be observed that at lower SNR values, the difference in maximum eigenvalue of signal plus correlated noise case
and correlated noise only case becomes very small and it becomes difficult to distinguish signal from noise and furthermore, this effect becomes more prominent at higher values of SCN.

Figure 7 shows the normalized MSE versus SNR plot for signal plus correlated noise case for different SCNs of the noise correlation matrix. From the figure, it can be observed that for all values of SCN, normalized MSE decreases with the SNR. We can estimate the SNR with 10 \% normalized MSE upto -1 dB and after 3 dB, we can estimate with normalized MSE of 1 \%. From the figure, it can be noted that the normalized MSE becomes almost stable if we go to higher SNR values beyond 3 dB and it increases for higher values of SCN at lower SNR values.

\subsection{FS Operating Point}
Figure 8 shows the ratio of correct sensing versus FS rate ($M$) for $\varepsilon=3.5$. The FS rate has been increased from $1$ to $11$ and noise correlation has been calculated using eqn. (\ref{eq: linear}) for different values of $M$. It can be noted that the sensing performance increases with the FS rate for white noise scenario. However, at the same time, noise becomes correlated due to FS and increasing the sampling rate does not monotonically increase the performance. From Fig. 8, it can be noted that for $N$=$60$, SNR=-$5$ dB, the performance increases up to FS rate $M=8$ and for $M>8$, the sensing with
eqn. (\ref{eq: decmp1}) saturates. It can be observed that increasing sampling rate enhances the sensing performance up to a certain FS rate, however, this also increases the complexity in the receiver. Thus it can be concluded that optimum sampling rate should be chosen at the receiver without increasing further complexity since larger rate does not enhance the performance due to noise correlation.

\section{Conclusion}
In this paper, the performance of eigenvalue based sensing has been analyzed in presence of noise correlation. This case often appears due to imperfections in filtering or oversampling and results in non-Wishart covariance matrices. A new SCN-based threshold has been proposed for improved sensing in the presence of noise correlation. Furthermore, an SNR estimation technique based on the maximum eigenvalue of received
signal's covariance matrix has been proposed and the performance of proposed technique has been analyzed with normalized MSE. It has been shown that SNRs upto 0 dB can be reliably estimated without any knowledge of noise variance. Moreover, the performance of FS based SS technique is studied and it has been noted that
SS efficiency increases with FS rate up to a certain limit and it does not provide performance advantage beyond this limit. Therefore, it can be concluded that an optimal operating point for the FS rate should be selected to maintain a good trade-off between performance and complexity.
\begin{center}
   APPENDIX \\
Random Matrix Theory Preliminaries
\end{center}
Let $F_{\mathbf X}(x)$ be the eigenvalue probability density function of a matrix $\mathbf X$.

\begin{thm}
The Stieltjes transform $\mathcal S_{\mathbf X}(z)$ of a positive semidefinite matrix $\mathbf X$ is defined by \cite{Ant:04}:
\begin{equation}\label{}
  \mathcal {S}_{ \mathbf X}(z)= \mathbb E \left[ \frac{1} {\mathbf X-z} \right]=\int_{- \infty}^{\infty}  {\frac{1}{\lambda-z} dF_{\mathbf X} (\lambda)}
 \end{equation}
\end{thm}

The a.e.p.d.f. of $\mathbf X$ is obtained by determining the imaginary part of the Stieltjes transform $\mathcal {S}_{\mathbf X}$ for real arguments:
\begin{equation}\label{}
   \lim_{y \to 0^+}  \frac{1}{\pi} \mathrm{Im} \{ \mathcal S_{\mathbf X} (x+jy) \}
  \end{equation}

\begin{thm}
The R transform is related the inverse of Stieltjes transform by the relation \cite{Ant:04}:
\begin{equation}\label{}
  \mathcal{R}_{\mathbf X}(z)= \mathcal{S}_{\mathbf X}^{-1}(-z)-\frac{1}{z}
   \label{eq: Sttransform}
\end{equation}

\end{thm}
\begin{thm}
For a Wishart random matrix $\mathbf X$, the R transform of the density of eigenvalues of $\mathbf X \mathbf X^H$ is defined as\cite{Ant:04}:
\begin{equation}\label{}
\mathcal {R}_{\mathbf X} (z)=\frac{\beta}{1- z}
\label {eq: Rmp}
\end{equation}
For any $a>0$,
\begin{equation}\label{}
  \mathcal{R}_{a  \mathbf X}=a \mathcal{R}_{\mathbf X}(az)
  \label {eq: Rmpp}
\end{equation}
\end{thm}

\begin{thm}
For a Wishart random matrix $\mathbf X$, the $\Sigma$ transform of the density of eigenvalues of $\mathbf X \mathbf X^H$ is defined as\cite{Ant:04}:
\begin{equation}\label{}
  \Sigma_{\mathbf X}(z)=\frac{1}{z+\beta}
  \label {eq: Str}
\end{equation}
\end{thm}

\vspace{15 pt}
\begin{center}
  ACKNOWLEDGEMENT
\end{center}
This work was supported by the National Research Fund, Luxembourg under AFR (Aids Training-Research) grant for PhD project on ``Spectrum Sensing, Resource Allocation and Resource Management Strategies for Satellite Cognitive Communications'', under the CORE project ``CO2SAT: Cooperative and Cognitive Architectures for Satellite Networks''.
\newpage
\bibliographystyle{IEEEtran}
\bibliography{rmtjournal}

\newpage

\begin{figure}[!h]
\begin{center}
\includegraphics[width=3.6 in]{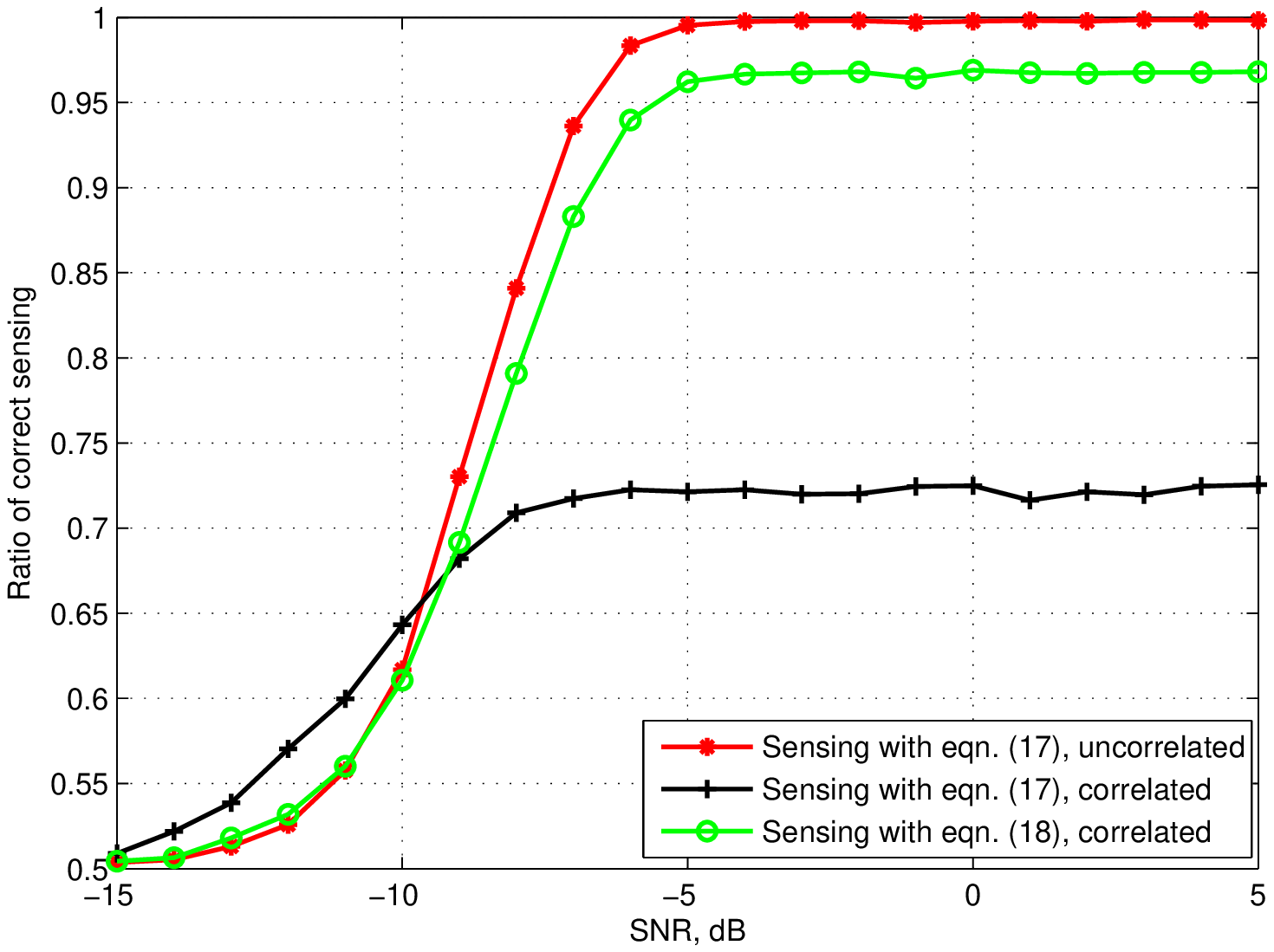}
\caption{\footnotesize{Sensing performance versus SNR with eqn. (27) and (28) ($\beta=1/6, \rho=0.5$)}}\label{fig_1}
\end{center}
\end{figure}

\vspace{60 pt}

 \begin{figure}[!h]
\begin{center}
\includegraphics[width=3.6 in]{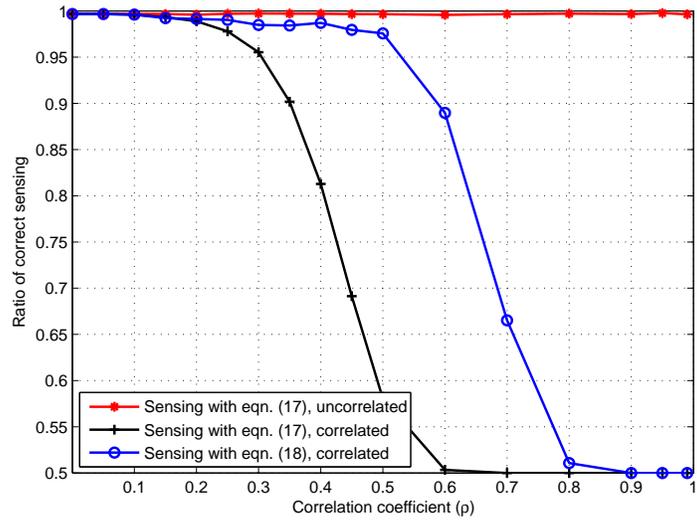}
\caption{\footnotesize{Sensing performance versus correlation coefficient in presence of white and correlated noise (SNR$=-6$ dB, $\beta=1/6$)}}\label{fig_1}
\end{center}
\end{figure}

\newpage

%

\begin{figure} [!h]
\begin{center}
\includegraphics[width=3.2in]{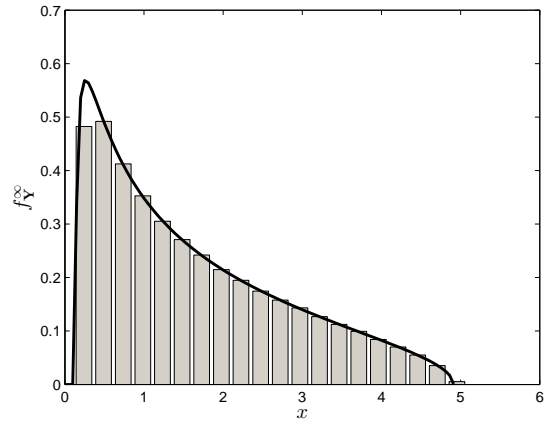}
\caption{\footnotesize{Theoretical and simulated eigenvalue distribution of signal plus white noise for SNR$=-2$ dB, $\beta$=1, $N=50$}}
\end{center}
\end{figure}

\vspace{70 pt}

\begin{figure}[!h]
\begin{center}
\includegraphics[width=3.2 in]{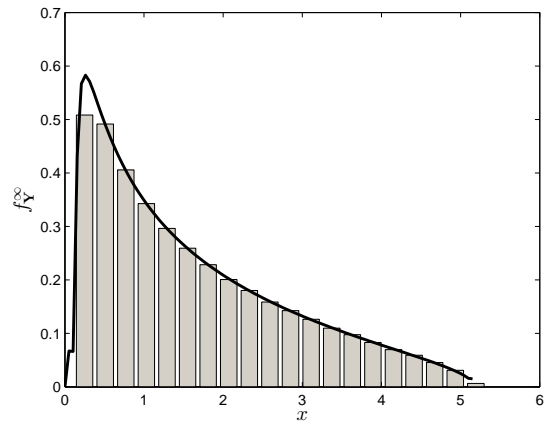}
\caption{\footnotesize{Theoretical and simulated eigenvalue distribution of signal plus correlated noise for SNR$=-2$ dB, $\beta=1$, SCN$=3$, $N=50$}}
\end{center}
\end{figure}

\newpage

\begin{table} [!h]
\caption{Look up table for different values of condition numbers}
\centering
\begin{tabular}{l|c|c|c|c|c}
 SCN &  $\beta$ & SNR (dB) & $\lambda_{max}$  & $\lambda_{max}$  & $\lambda_{max}$   \\
  &    &  & Signal plus correlated Noise & correlated Noise & Signal plus white Noise  \\\hline
  1.5& 1  & 5 & 14.21 & 4.07 & 14.20  \\
  1.5& 1  & 2 & 7.83 & 4.10  & 7.81   \\
  1.5& 1  & 0 & 5.95  & 4.13 & 5.91 \\
  1.5& 1  & -2 & 4.96 & 4.09 & 4.93  \\
  1.5& 1  & -4 & 4.61  & 4.06  & 4.52   \\
  1.5& 1  & -6  & 4.34  & 4.08  & 4.29  \\
  1.5& 1  & -8  & 4.28 & 4.11 &  4.17  \\
  1.5& 1  & -10  & 4.29 & 4.07 &  4.23  \\ \hline
  2& 1  & 5 &  13.98 & 4.17 & 13.98  \\
  2& 1  & 2 & 7.77  & 4.20  &  7.76  \\
  2& 1  & 0 & 5.93  & 4.18 & 5.85  \\
  2& 1  & -2 & 5.04  & 4.16 & 4.91  \\
  2& 1  & -4 & 4.68  & 4.21  & 4.52   \\
  2& 1  & -6  & 4.44  & 4.16  & 4.29  \\
  2& 1  & -8  & 4.37  & 4.17 & 4.20   \\
  2& 1  & -10  & 4.31 & 4.21 & 4.17   \\\hline
  2.5 & 1  & 5 & 13.97 & 4.29 & 13.96  \\
  2.5 & 1  & 2 & 7.91   & 4.23 & 7.85  \\
  2.5 & 1  & 0 &  5.95 & 4.23  & 5.86 \\
  2.5 & 1  & -2 & 5.15  & 4.26 &  4.88\\
  2.5 & 1  & -4 &  4.87 & 4.24   & 4.52  \\
  2.5 & 1  & -6  & 4.54  & 4.30  & 4.31\\
  2.5 & 1  & -8  & 4.51  &  4.25 & 4.21 \\
  2.5 & 1  & -10  & 4.34  & 4.23 & 4.15 \\ \hline
 \end{tabular}
\label{tab: practicalterr}
\end{table}


\vspace{30 pt}

\begin{figure}[!h]
\begin{center}
\includegraphics[width=3.6in]{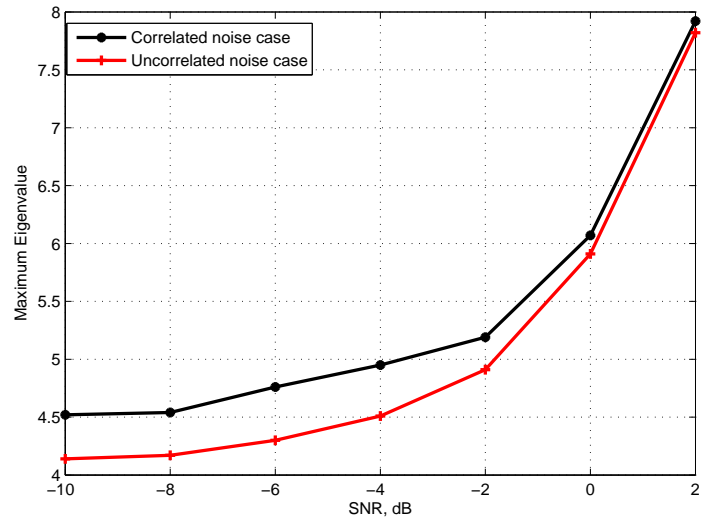}
\caption{\footnotesize{Maximum eigenvalue versus PU SNR for correlated and white noise case (SCN$=3$, $\beta=1$)}}
\end{center}
\end{figure}

\newpage

\begin{figure} [!h]
\begin{center}
\includegraphics[width=3.6in]{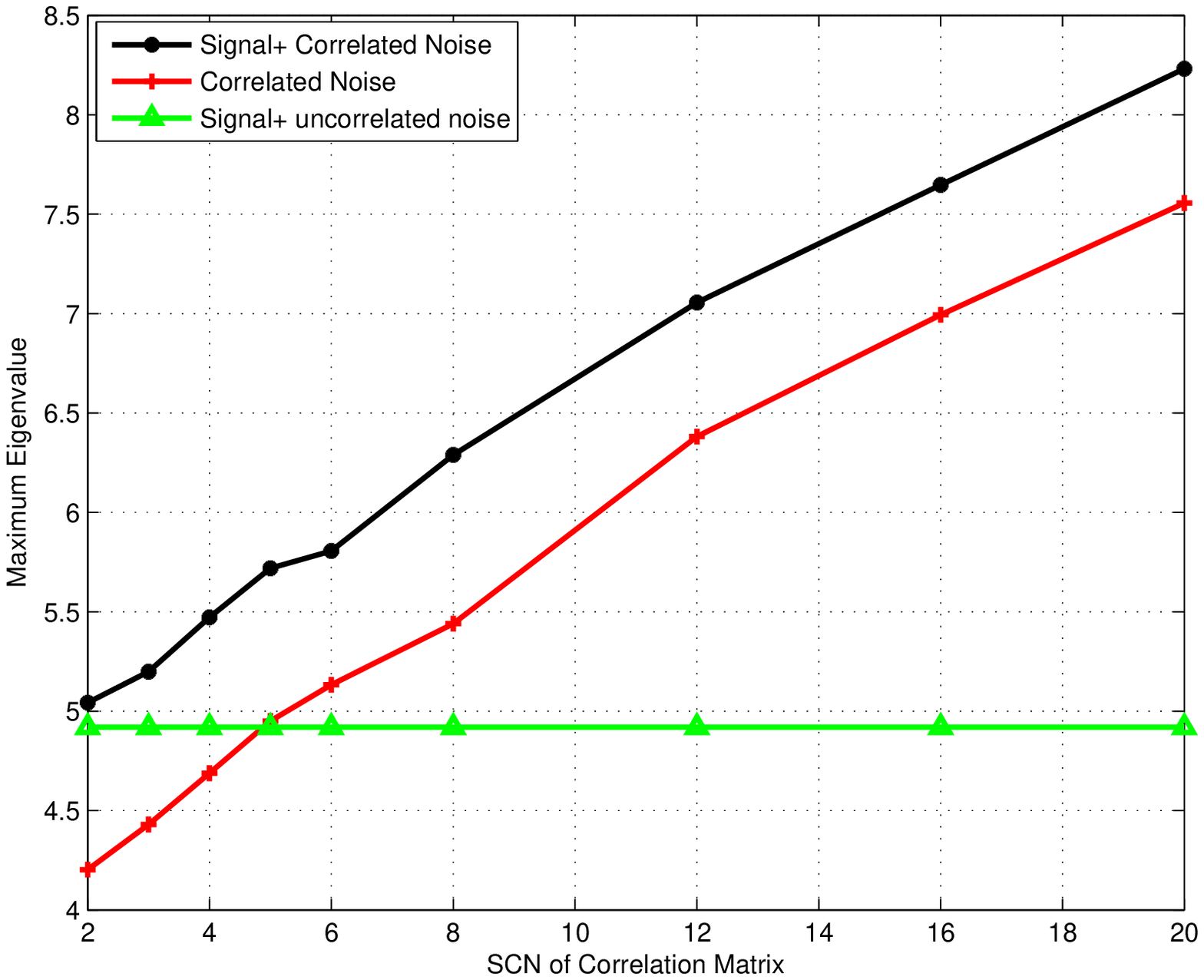}
\caption{\footnotesize{Maximum eigenvalue versus SCN for correlated and white noise case (SNR$=-2$ dB, $\beta=1$)}}
\end{center}
\end{figure}


\vspace{70 pt}


\begin{figure} [!h]
\begin{center}
\includegraphics[width=3.6in]{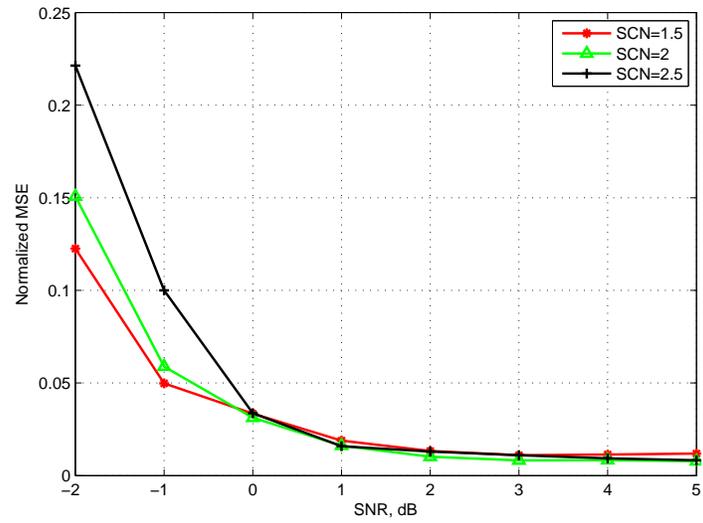}
\caption{\footnotesize{ Normalized MSE versus SNR for SNR estimation using proposed technique}}
\end{center}
\end{figure}


\begin{figure} [!h]
\begin{center}
\includegraphics[width=3.6 in]{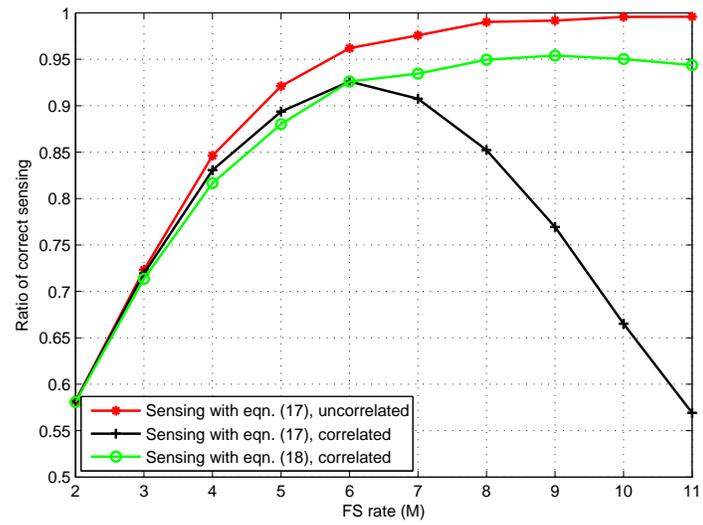}
\caption{\footnotesize{Sensing performance versus FS rate ($N=60$, SNR$=-5$ dB)}}\label{fig_1}
\end{center}
\vspace{-26 pt}
\end{figure}

\end{document}